\documentstyle[twocolumn,aps]{revtex}
\begin{document}

\draft

\title{Composite fermions traversing a potential barrier}

\author{L. Brey$^1$ and C. Tejedor$^2$}
\address{$^1$Instituto de Ciencia de Materiales (CSIC).
Universidad Aut\'onoma de Madrid C-12,
Cantoblanco, 28049, Madrid, Spain.  }
\address{$^2$Departamento de Fisica de la Materia Condensada.
Universidad Aut\'onoma de Madrid,
Cantoblanco, 28049, Madrid, Spain.  }

\date{\today}

\maketitle

\begin{abstract}
Using a composite fermion picture,
we study the lateral transport between two two-dimensional electron gases,
at filling factor 1/2, separated by a potential barrier.
In the mean field approximation, composite
fermions far from the barrier do not feel a magnetic field while in the
barrier region the effective magnetic field is different from zero.
This produces a cutoff in the conductance when represented
as a function of the thickness and height of the barrier.
There is a range of barrier heights for which
an incompressible liquid, at $\nu =1/3$, exists in the barrier region.

\end{abstract}
PACS number 73.40.Hm


The presence of a high magnetic field $B$ perpendicular to a two dimensional
electron gas produces the Quantum Hall effect (QHE) characterized by
the appearance of plateaus at rational values of $e^2/h$ of
the off-diagonal conductance $\sigma _{xy}$ and zeros  in the diagonal
conductance $\sigma _{xx}$. QHE is usually analyzed
separating the set of features in two, depending on whether the plateau appears
at a filling factor $\nu _e=2\pi \ell ^2 n$ which is an integer number (IQHE),
or
a fractional one (FQHE). In this expression $\ell = \sqrt {\hbar c/eB}$
is the magnetic length and $n$ the two dimensional electron density. The reason
for this division is
the different physical origin of the energy gap between occupied and unoccupied
states of the system, which is responsible for the QHE.
In the case of IQHE the gap originates from the quantization of the kinetic
energy in the presence of $B$ while for the FQHE the gap is
due to the electron-electron interaction. As from the
experimental point of view both IQHE and FQHE present similar behavior,
it is appealing to look for a unified representation of the two
processes\cite{jain}. This is the goal of the composite fermions (CF) picture
of the FQHE. In this framework, the relevant quasiparticles
are CF built up as electrons attached to an even number,  $\tilde { \phi}$,
of flux quanta. Instead of working with the electronic Hamiltonian, a singular
gauge transformation is performed and the system is described by a Hamiltonian
for the CF which interact by both Coulomb and Chern-Simons terms\cite{alopez}.
Within a mean field approach, CF move in an effective magnetic field
$B_{eff} = B - \tilde { \phi}  n h c /e$.
We made the choice of $\tilde{\phi}$=2 so that
when $B$ is such that the electron filling factor is one half, CF do not
feel any effective magnetic field and present a Fermi surface\cite{HLR}.
In the case of Coulomb interaction between the electrons, the Fermi
surface does not dissapears when fluctuations are included\cite{HLR,nota}.
At the same time, an incompressible electron liquid in the FQHE regime given by
$\nu _e=p/(1+\tilde { \phi }p)$ corresponds to the IQHE regime for CF with
filling
factor $\nu _{CF}=|p|$\cite{jain,alopez,HLR,kim}. In other words,
a mean field aproach for CF gives an IQHE-like description for the FQHE
although their physical origins are different.
A strong support for the CF theory is that several predictions coming
from it have been experimentally confirmed\cite{exp}.

We study the transport of CF through a potential
barrier in the presence of an external magnetic field,
corresponding to an electronic filling factor $\nu _e =1/2$.
As charge is expelled from the barrier region,
the effective field $B_{eff}$ affecting the CF is not homogeneous anymore.
This produces two effects: i) the localized effective field deflects
the carriers in such a way that, if they do not
have enough kinetic energy, they are not able to pass through the {\it
magnetic barrier} and
ii) a region of incompressible liquid can appear at the barrier. In particular
we will show how an incompressible phase $\nu _e=1/3$, i.e. $\nu _{CF}=1$,
forms and separates the two phases of compresible liquid
$\nu _e=1/2$, i.e. CF with Fermi surface. This incompressible phase in the
barrier region controls the current in the linear regime because it is stable
in the presence of low density of impurities contrarily the {\it magnetic
barrier} effect which is quenched by impurity scattering.

 We consider that the potential barrier is along  $x$ direction and the
system is translationally invariant in the other one, $y$.
Therefore we describe the effective magnetic field by using the Landau gauge
$ {\bf A_{eff}}=(0,\int dxB_{eff}(x),0)$.
In order to compute conductances through the barrier, the required ingredient
the probability that one CF coming from the left part of the barrier passes
This must be computed taking into account the dependence of the effective
magnetic field on the charge density,
so selfconsistency between $B_{eff}(x)$ and $n(x)$ is required.
This task is accomplished in the following way.
We consider an external potential
$V_0(x)$ which, after the charge rearrangement, will give a
barrier for the CF of the form showed in Fig.1a;
$V(x)= V_b \left [ \Theta ( x + d_b /2)- \Theta(x - d_b /2) \right ] $.
For this given final potential it is necessary to know the amount of charge
and the effective magnetic field in the barrier region. We obtain
these quantities self-consistently in the Hartree
approximation \cite{brey}.
In this approach the Schr\"{o}dinger equation for CF\cite{brey} takes the
form,
\begin{eqnarray}
\left \{
{{\hbar \omega _c } \over 2} \ell ^ 2 \left [ -  {{d^2} \over {dx^2}}
+ \left (k_y+ { e \over c} { {A _ {eff} (x) } \over {\hbar}} \right )^2
\right ]+ V(x) \right \} \\ \nonumber
{{e^ { i k_y y }} \over { L _y }}\phi_{k_y, \varepsilon} (x) =
\varepsilon {{e^ { i k_y y }} \over { L _y }}\phi _{k _y , \varepsilon} (x)
\end{eqnarray}
here $\hbar k _ y $ is the momentum of the CF in the $y$ direction, $L_y $
is the sample length in the $y$-direction  and
$\omega _c =eB/m c$ is the cyclotron frequency, $m$ being the renormalized
mass of the CF. Note  that in this expresion $V(x)$
is the final potential seen by the CF, i.e. the sum of
the external potential, the Hartree potential and the potential
induced by the motion of the magnetic flux tubes\cite{brey,Fetter}.
The effective vector potential is
\begin{equation}
A_{eff}(x) =  { { c \hbar } \over {e}}  \left ( { {x} \over {\ell ^ 2}}
- \tilde {\phi} \pi  \int \! d x' n (x') {\rm sgn} (x -x')  \right )
 \, \, \,
\end{equation}
where the density $n(x) = \sum _{ k _y, \varepsilon } | \phi _ {k _ y ,
\varepsilon }(x) | ^ 2/L_y$ is a sum restricted to energies lower that the
Fermi energy, which is $\hbar \omega _c /2$ for the case of electronic
filling factor $\nu _ e$=1/2 \cite{brey}.
By solving Eq.1-2, we get the self-consistent $n(x)$ and $B_{eff} ( x) $.
The amount of charge in the barrier region is obtained by integrating $n(x)$
in a region of width $D$. This is a cut-off, convenient for numerical
calculations, large
enough to contain all the effects coming from charge rearrangements.
The amount of charge in the barrier region is represented  by the average
electron filling factor at the barrier $\nu _e^b$.
It must be obviously smaller than $\nu _e = 1/2$ and one can expect that the
system tends to condensate in this region at values of the density where
an incompressible liquid appears. We will see below that this is
precisely the result we get.

 Once we know the self-consistent effective magnetic field in the barrier
region we are in condition to compute the transmission coefficient
through the barrier. Since in the barrier region $B_{eff} \neq 0$, there is
a step, $\Delta A = A_{eff} ( -\infty)-A_{eff} ( \infty)$,
in the effective vector potential (see Fig.1(c)). This step
produces that the $y$-component
of the canonical momentum of the CF changes by ${ e \over c} \Delta A$, in
passing from a point to the left of the barrier, $-D/2$, to a point to the
right, $D/2$, which are not affected by the existence of the barrier.
Therefore, in order to calculate the transmission coefficient for a CF with
mechanical momentum  $\hbar (k_x,k_y)$ and energy $\varepsilon = \hbar ^ 2
(k_x^2+k_y^2)/2m$, we  match the two independent solutions
of Eq.1 at energy  $\varepsilon$,  with the planewave
$ \left ( e ^ {i k _ x x } + \rho e ^ {-i k _ x x} \right ) $
at $ x=- D/2$ and with $ \tau e ^ {i k ^{'} _ x x } $
at $x=D/2$. Here $ k _ x ^ {'} = \left ( k _ x ^ 2 - ( {e \over {c \hbar }}
\Delta A ) ^ 2 + 2 k _ y {e \over {c\hbar}} \Delta A  \right ) ^ { 1/2} $,
as obtained from the conservation of both $k_y$ and energy.
{}From the matching of these wavefunctions the desired transmission coefficient
$T( \varepsilon, k _ y)= |\tau| ^ 2 {{ k _ x ^ {'}} / { k _ x}} $ is obtained.

 The first interesting property in the problem comes from the expression
for $k_ x ^ {'}$: Only CF with incident mechanical momentum
verifying the relation,
$ k _ x ^ 2 - ( {e \over { c \hbar }} \Delta A ) ^ 2 + 2 k _ y  { e \over {c
\hbar }} \Delta A > 0 $ ,
can pass through the barrier. This restriction
comes from the conservation of the mechanical momentum and the
energy, and is very easily understood in classical terms. A charged
incident particle with momentum $ {\bf p}= ( p_x, p _y)$, is deflected
when arrives to a spatial region, $-D/2<x<D/2$, where a magnetic field
$B$ exists. Only when the incident momentum is high enough,
the particle is able to cross the finite $B$ region.
Therefore, there is a region in momentum space, shown in Fig.1d, where
the transmission coefficient through the
barrier is zero. In the other regions in momentum space,
the transmission coefficient can take finite values.
This condition gives us a cutoff for the conductance
of the CF system in the presence of a barrier.

The conductance is given by an integral of the transmission coefficients
over all the incident states with energy $ E_F$\cite{peeters},
\begin{equation}
G= G_0
\int _{-\pi /2}^{\pi /2} T(E_F,\sqrt {2E_F}\sin \theta) \cos \theta d\theta \,
\, \, ,
\end{equation}
with $ G _ 0 = \frac {e^2 k_F }{\hbar } L _ y$,
$k_F$ being the Fermi wavevector.
This conductance for non-interacting CF, treated in mean field approximation,
corresponds to the conductance of the 2D electron gas at $\nu =1/2$\cite{HLR}.
In the incident mechanical momentum space, states at the Fermi energy are on a
semicircle centered at $(0,0)$ (see Fig.1d). If this semicircle is contained
in the region of $T=0$ the  conductance is zero. In order
to have $G\ne0$ the Fermi line must cross the boundary separating
the region of $T=0$ from that of $T\ne0$. This is only possible if the
barrier is narrow or low enough to allow a large charge density
in the barrier region, implying $B_{eff}$ is small there.

It must be stressed that, in the description we are using, the CF is
always constituted by one electron carrying two flux quanta. Therefore,
the scheme can not be directly applied to tunneling processes (i.e. $E_F<V_b$)
in which the density at the barrier is practically zero. In such a case the
CF at the barrier region could be strongly affected in its
constitution.

The numerical results for $G$ and $\nu _e^b$ as a function of barrier height
are presented in Fig.2 and Fig.3 for barriers thickness of
$4\ell$ and $16\ell$, respectively. In order to see more clearly the effect
of $B_{eff}$ on the CF, the same magnitudes are plotted
in dashed lines for electrons passing through a barrier without any
magnetic field. In this case, both $G$ and $\nu _e ^ b$ decrease
monotonously with incresing barrier height. When $V_b=E_F$ there is
very small amount of charge at the barrier region and the conductance
is enormously reduced because tunneling is the only mechanism for
transport. The results are quite different for CF, in the metallic regimen
($\nu _e=1/2$), traversing a region with a potential barrier.
The effect of the cutoff caused by $\Delta A$ on the
transmission coefficient, produces the quenching of $G$ for barriers
significantly lower than $E_F$. This is clear by noting that in the case of
CF in the metallic regime, and supposing an abrupt {\it magnetic barrier},
the cutoff for $T\neq0$ at $E_F$ is given by $ ( 1 - 2 \nu _ e ^ b) d_b / \ell
= 2 k _F \ell$. $\nu _e=1/2$ implies $k_F$=$\ell ^{-1}$\cite{brey} and we
obtain
 that the critical $\nu _ e ^ b$ for the barrier $4\ell$ and $16\ell$ thick
are 0.25 and 0.469 respectively. From  comparing Fig.2a (Fig.3a)  with
Fig.2b (Fig.3b), we see that the quenching of $G$ occurs roughly at these
values of $\nu _e ^ b$.

The second, and even more important effect is observed in the electron
filling factor at the barrier. This magnitude starts decreasing monotonously
and remains finite even for barrier height at which $G$=0. But
further increasing barriers, $\nu _e ^b$ shows an abrupt decrease for a
critical value of $V_b$. For $d_b=16\ell$ a plateau with value $\nu _e^b
=1/3$ develops before the abrupt decrease.
The physical origin of this behaviour can be understood by looking at the
schematic band structure of the system shown in Fig.4
for three increasing values of the barrier height and, consequently for
increrasing effective magnetic field. The dispersion relation consists
basically on two  filled parabolas, with the bottoms
shifted by the quantity $ {e\over {c\hbar}} \Delta A$. Each of
these parabolas corresponds to CF located at the left and the right side of the
barrier. Due to the existence of  $B_{eff}\neq$0 in the barrier region, some
Landau-like bands (LLB) appears. For low barriers, i.e. low $B_{eff}$ and
low $\Delta A$, the parabolas cross each other at low energy and the LLB are
placed in the continuum defined by the parabolas and there are not localized
states at the barrier region (Fig.4a).
For increasing barrier, the shift between the bottom of the parabolas
increases and a set of LLB appears in the region between parabolas (Fig.4b).
When only the lowest energy  LLB is occupied, an
incompressible liquid, with $\nu _{CF}=1$, exists in the barrier region.
Therefore the system is stable for
$\nu _e^b=1/3$ giving the plateau observed in Fig.2 and Fig.3.
As the LLB are dispersionless in the barrier region,
the plateau  at $\nu _ e ^b$=1/3  disappears suddenly when, for higher
barriers,
$B_{eff}$ is big enough so that the lowest energy LLB is
empty and there are not localized occupied states at the barrier (Fig.4c).
The plateau in the filling
factor at the barrier is then the signature of the existence of an
incompressible liquid at the barrier separating the two metallic regions
at the left and the right.

An important issue is whether the two effects found here, the cutoff for $G$
and the incompressible region at the barrier, are robust with respect to
the presence of a small random potential due to impurities
or imperfections in the system. The existence of a random potential
would relax the momentum conservation which is in the
origin of the existence of the cutoff for $G$. Therefore the first consequence
of
having impurities is that the current would not be quenched as before.
However the incompressible liquid is rather robust, with respect to the
presence of a weak random potential, provided that
the energy gap in the barrier region persists in the presence of impurities.
Therefore, we expect that the imcompressible region would persist
in samples in which the mobility is high enough so that the $\nu_ e $=1/3 FQHE
can be observed. The existence of an incompressible liquid in the barrier
region (no states at $E_F$) produces that $G$ would be zero for barriers  in
which $\nu _e^b$ presents a plateau. Therefore,
the quenching of the conductance persist although for barriers higher
than before and due to a different physical origin.

We are indebted to L.Martin-Moreno for fruitful discussions.
The help received from J.J. Dorado is deeply acknowledged.
This work has been supported in part by the Comisi\'on Interministerial
de Ciencia y Tecnologia of Spain under contract No. MAT 94-0982-C02.

\begin{figure}
\caption {
Schematic plot of (a) the potential barrier $V(x)$,
(b) charge density profile $n(x)$, and (c) effective vector
potential $A_{eff}(x)$. In (d),
the separation between the regions with $T=0$ and with $T\neq0$
is plotted, in the space of the mechanical momentum of the incident CF.
The semicircles are the Fermi lines in the case
in which $G=0$ (dashed line) and in the case in which $G$ can be different
from zero (dotted line).}
\label{fig1}
\end{figure}
\begin{figure}
\caption {
Variation of the filling factor in the barrier region, $\nu _ e ^ b$ (a)
and the conductance $G$ (b), as a function of the barrier height
$V_b$ in units of $\hbar \omega _c$, for a barrier 4$\ell$ thick.
The dashed lines correspond to the case of electrons without magnetic field. }
\label{fig2}
\end{figure}
\begin{figure}
\caption {
Variation of the filling factor in the barrier region, $\nu _ e ^ b$ (a)
and the conductance $G$ (b), as a function of the barrier height
$V_b$ in units of $\hbar \omega _c$, for a barrier 16$\ell$ thick.
The dashed lines correspond to the case of electrons without magnetic field. }
\label{fig3}
\end{figure}
\begin{figure}
\caption { Schematic representation of the band structure of CF
in the metallic regime and in presence of a potential barrier for
different heights of the potential barrier. }
\label{fig4}
\end{figure}


\begin{references}
\bibitem{nota}We treat CF as having a renormalized finite effective mass
although this is still a matter of controversy and such a mass has not been
measured for $\nu _{e}=1/2$\cite{exp}.
\end{references}
\end{document}